\documentclass[twocolumn]{autart}    

\usepackage{ifpdf}
\usepackage{amsmath}
\usepackage{amsfonts}
\usepackage{mathrsfs}
\usepackage{color}
\usepackage{graphicx}
\usepackage{subfigure}
\usepackage{epstopdf}
\usepackage{lineno}
\usepackage{cite}
\usepackage{enumerate}
\usepackage{algorithm, algorithmic}
\usepackage{eqnarray}
\usepackage{bm}
\usepackage{wasysym}%
 \usepackage{booktabs} 
\usepackage{tabularx}

\newtheorem{theorem}{\textbf{Theorem}}
\newtheorem{lemma}{\textbf{Lemma}}

\newtheorem{definition}{\textbf{Definition}}
\newtheorem{example}{\textbf{Example}}

\newtheorem{remark}{Remark}
\newtheorem{assumption}{Assumption}

 \newcommand{\cvar}{\text{CVaR}}

\newcommand{\DC}{\mathcal{D}}

\newcommand{\PC}{\mathcal{P}}

\newcommand{\RC}{\mathcal{R}}

\newcommand{\TC}{\mathcal{T}}

\newcommand{\XC}{\mathcal{X}}

\DeclareMathOperator*{\argmin}{arg\,min}
\usepackage{graphicx}          

\begin{document}

\begin{frontmatter}

\title{Risk-Averse Learning with Non-Stationary Distributions\thanksref{footnoteinfo}} 

\thanks[footnoteinfo]{This work was supported  by the European Research Council
(ERC) Consolidator Grant ”Safe data-driven control for humancentric systems (CO-MAN)” under grant agreement number 864686, by the Swedish Research Council Distinguished Professor Grant 2017-01078, Knut and Alice Wallenberg Foundation, Wallenberg Scholar Grant, the Swedish Strategic Research Foundation CLAS Grant RIT17-0046, AFOSR under award \#FA9550-19-1-0169, and  NSF under award CNS-1932011. Corresponding author: Zifan Wang.}

\author[ITR]{Siyi Wang}\ead{siyi.wang@tum.de},   
\author[DCS]{Zifan Wang}\ead{zifanw@kth.se},                
\author[LIDS]{Xinlei Yi}\ead{xinleiyi@mit.edu},  
\author[DUKE]{Michael M. Zavlanos}\ead{michael.zavlanos@duke.edu},   
\author[DCS]{Karl H. Johansson}\ead{kallej@kth.se}, 
\author[ITR]{Sandra Hirche}\ead{hirche@tum.edu}

\address[ITR]{Chair of Information-oriented Control (ITR), Department of Electrical and Computer Engineering, Technical University of
Munich, 80333 Munich, Germany
} 
\address[DCS]{Division of Decision and Control Systems, School of Electrical Engineering and Computer Science, KTH Royal Institute of Technology,
10044  Stockholm, Sweden}    
\address[LIDS]{Lab for Information \& Decision Systems, Massachusetts Institute of Technology, Cambridge, MA 02139, USA} 
\address[DUKE]{Mechanical Engineering and Material Science, Duke University, Durham, NC 27708, USA} 

\begin{abstract}                     

Considering non-stationary environments in online optimization enables decision-maker to effectively adapt to changes and improve its performance over time. In such cases, it is favorable to adopt a strategy that minimizes the negative impact of change to avoid potentially risky situations. In this paper, we investigate risk-averse online optimization where the distribution of the random cost changes over time. We minimize risk-averse objective function using the Conditional Value at Risk (CVaR) as risk measure. Due to the difficulty in obtaining the exact CVaR gradient, we employ a zeroth-order optimization approach that queries the cost function values multiple times at each iteration and estimates the CVaR gradient using the sampled values. To facilitate the regret analysis, we use a variation metric based on Wasserstein distance to capture time-varying distributions. 
Given that the distribution variation is sub-linear in the total number of episodes, we show that our designed learning algorithm achieves sub-linear dynamic regret with high probability for both convex and strongly convex functions. Moreover, theoretical results suggest that increasing the number of samples leads to a reduction in the dynamic regret bounds until the sampling number reaches a specific limit. Finally, we provide numerical experiments of dynamic pricing in a parking lot to illustrate the efficacy of the designed algorithm. 
\end{abstract}
\end{frontmatter}

\section{Introduction}\label{sec:introduction}
Online convex optimization is a powerful framework that deals with decision-making problems in dynamic and uncertain environments \cite{hazan2006efficient}. 
It has many applications, including traffic routing \cite{sessa2019no}, resource allocation \cite{chen2017online}, and online marketing \cite{gordon2008no}.
In online optimization, the decision maker sequentially updates its decision in a changing environment  relying on historical information such as observations of previous actions and costs. 
The decisions generated by the optimization algorithm induce a sequence of associated cost values.  
The performance of the algorithm is evaluated using the notion of regret \cite{hazan2016introduction}, which is the accumulated loss generated by the algorithm against the optimal actions in hindsight.

Non-stationary environments describe scenarios where the underlying conditions of the system change over time. 
The reason for environmental changes can be variations in the distribution of the stochastic cost function. 
%
For instance, in dynamic pricing for vehicle parking \cite{ray2022decision}, the pricing depends on real-time changes in demand and supply; conversely, price adjustments influence the distribution of the occupancy rate.
In non-cooperative games, 
the objective function of each agent follows a distribution, which may evolve over time in response to action updates of other agents \cite{narang2022learning}. 
%
A further example is performative prediction, in which the data distribution evolves with the decisions over time. It has gained significant attention in the machine learning community recently  \cite{perdomo2020performative,miller2021outside}. 

Compared with the standard regret analysis assuming a stationary environment, dynamic regret provides a more relevant performance evaluation by considering the impact of fluctuations in non-stationary environments \cite{besbes2015non,zhao2018data}. 
Variation metrics are introduced to analyze the dynamic regret, for example, the variations in the cost functions \cite{besbes2015non} and the variations in the optimal actions \cite{zhao2021bandit}, which is also known as the path length of the comparators. When making decisions under uncertainty, it is often essential to consider the entire distribution of potential outcomes rather than focusing on specific optimal points.
Specifically, \cite{jiang2020online,shames2020online} use the Wasserstein distance metric, which defines the dissimilarity between probability distributions, to quantify the changes of non-stationary distributions. 
In this paper, we employ the distribution variation metric proposed in \cite{jiang2020online}.
 
When the decision-maker is sensitive to potential negative consequences, 
its primary consideration is not 
minimizing the expected cost, but rather reducing the risk of a catastrophe.
For example, in the financial market, it is unfavorable to pursue a strategy that entails high risks despite offering the highest expected reward.
%
Some measures are proposed to model the potential risk, such as Value at Risk (VaR) \cite{linsmeier2000value} and Conditional Value at Risk (CVaR) \cite{rockafellar2002conditional}. Given a risk level $\alpha \in (0,1]$, the CVaR value describes the average value of the $\alpha$-tail distribution of the stochastic cost.
It has a coherent risk measure property, which offers some mathematical properties that facilitate theoretical analysis. 
The classical paper \cite{rockafellar2000optimization} formulates the computation of the CVaR value as an optimization problem by introducing an additional decision variable to construct an augmented objective function. 
It enables the application of CVaR for the optimization problem in bandit optimization \cite{cardoso2019risk}, online games \cite{wang2022zeroth,wang2022risk} and safe control \cite{chapman2021toward,kishida2022risk,chapman2019risk}.
However, since the computation of the CVaR gradient relies on the distribution of the stochastic cost, CVaR optimization problems rarely enjoy a closed-form expression. To handle this problem, a common approach is to estimate the CVaR gradient using zeroth-order optimization algorithms, see \cite{cardoso2019risk,wang2022risk}.


\subsection{Our Contributions}
In this paper, we investigate online CVaR optimization, when the distribution of the stochastic cost changes over time.
To the best of our knowledge, such risk-averse learning with non-stationary distributions has not been explored in the literature.
As mentioned above, the exact CVaR gradient is generally unknown. Hence,  we use a zeroth-order optimization algorithm to estimate it.
However, it is not possible to efficiently estimate the CVaR gradient with only a single sample of the random cost per iteration, especially when the distribution of the cost changes over time.
To address this issue, 
we propose a sampling strategy motivated by \cite{wang2022zeroth}, which queries function values multiple times at each iteration. 
Then we use the sampled function values to construct the empirical distribution function of the random cost and estimate the CVaR values. Based on these CVaR values, we use the zeroth-order optimization approach to construct the CVaR gradient estimate.

Additionally, we introduce the concept of distribution variation based on the Wasserstein distance metric to measure the variation of the non-stationary distributions.  
To ensure that the decision-maker is able to adapt to the changing distributions, the learning algorithm is periodically restarted. 
Then we analyze the dynamic regret of the designed risk-averse learning algorithm in terms of the distribution variation metric for both convex and strongly convex cost functions. 
Under mild condition on the learning rate and the smoothing parameter, the regret upperbound of the algorithm is minimized.
We show that the algorithm achieves sub-linear dynamic regret with high probability,  given that the distribution variation is sub-linear in the total number of iterations.
A tuning parameter $a>0$ regulates the number of samples, where a higher value of $a$ indicates a larger number of samples. 
Our results suggest that the regret bound achieved by the algorithm decreases with the increasing sampling number until it reaches a certain limit. Denote $V_D$ as the distribution variation budget.
Table~\ref{table:1} summarizes the dynamic regret bounds for the convex and strongly convex cases with various values of $a$.  
It can be observed that when $0<a \le 1$, the strongly convex problem has the same regret bound as the convex problem. When $a > 1$, the regret bound of the strongly convex problem is strictly lower than that of the convex problem.
Finally, we illustrate our algorithm using the example of dynamic pricing in a parking lot. 
\begin{table}[h!]
\centering{\caption{Order of Regret}}
\begin{tabularx}{0.48\textwidth}{ 
   >{\centering\arraybackslash}X 
    >{\centering\arraybackslash}X 
   >{\centering\arraybackslash}X  } 
\toprule[1pt]
 Sampling parameter & Convex  &Strongly convex\\
\midrule[1pt]  
$0<a\le1 $ &  $T^{\frac{4}{4+a}}V_D^{\frac{a}{4+a}}$   & $T^{\frac{4}{4+a}}V_D^{\frac{a}{4+a}}$ \\ 
\hline 
$1<a \le \frac{4}{3}$ &  $T^{\frac{4}{5}}V_D^{\frac{1}{5}}$   &  $T^{\frac{4}{4+a}}V_D^{\frac{a}{4+a}}$  \\ 
\hline 
$a>\frac{4}{3} $ & $T^{\frac{4}{5}}V_D^{\frac{1}{5}}$   & $T^{\frac{3}{4}}V_D^{\frac{1}{4}}$   \\ 
\bottomrule[1pt]
\end{tabularx}\label{table:1}
\end{table}

\subsection{Related Works}
There is a growing interest in non-stationary online convex optimization, see \cite{besbes2015non,zhao2018data,ray2022decision}.
For example, \cite{besbes2015non} investigates non-stationary stochastic optimization for both convex and strongly convex functions, and analyzes the dynamic regret using the variations in the cost functions as the measure. In \cite{zhao2021bandit}, the authors investigate bandit convex optimization in non-stationary environments. 
To the best of our knowledge, only few works address environmental changes that results in non-stationary distributions, e.g., \cite{cao2020online,jiang2020online,shames2020online}. 
The authors in \cite{cao2020online} investigate online stochastic optimization with time-varying distributions using the variations of optimal points as the measure. 
Moreover, \cite{shames2020online} investigates online stochastic optimization in the strongly convex case with the variations of the optimal points, while the dynamic regret is not strictly sub-linear due to the fluctuation of gradient estimates. However, computing the distribution variations is generally more convenient than computing the variations in the optimal solutions, as the latter often necessitates multiple iterations of gradient descent updates.
In \cite{jiang2020online}, the dynamic regret of online constrained optimization is analyzed in terms of the Wasserstein-based non-stationarity budget (WBNB), which upperbounds the cumulative deviation of all the distributions from their average distribution. However, even when the distribution changes only once throughout all the iterations,  the WBNB can still be large and thus not applicable to our problem.

Another related research line is risk-averse learning using CVaR as the risk measure \cite{tamkin2019distributionally,soma2020statistical,urpi2021risk,cardoso2019risk,wang2022risk,wang2022zeroth}.
For instance, \cite{tamkin2019distributionally} proposes a risk-averse learning algorithm for the multi-armed bandit problem and provides an upper confidence bound of the CVaR value.
Moreover, \cite{wang2022risk} investigates the risk-averse learning algorithm in online convex games, which achieves sub-linear regret for each agent with high probability. Specifically, compared to \cite{wang2022risk}, where the sampling numbers are designed to decrease with time, we only require the total samples over all iterations to be over a certain threshold.
However, all the above works focus on the stationary environment, i.e., both the cost function and distribution do not change over time. With only a few exception, e.g., \cite{liang2021adaptive} explores the risk-averse online optimization in a non-stationary environment and in the context of the online portfolio selection problem with linear costs. This paper addresses a more generalize setting that can be applied to a broader range of applications.

\subsection{Outlines}
The remainder of this paper is structured as follows: Section \ref{sec:preliminaries} introduces preliminaries and problem formulation. Section \ref{sec:Main result} presents the main result on risk-averse learning under non-stationary distributions. Section \ref{sec:simulation} demonstrates the efficacy of the designed algorithm by numerical simulations. Section \ref{sec:conclusion} draws conclusions. \\
\textbf{Notations:} Let $\|\cdot\|$ denote the $l_2$ norm.  Let $\lceil \cdot \rceil$  denote the ceiling function.  Let $\mathbf{1}(\cdot)$ denote the indicator function. For a random variable $X$, let $X \sim \DC_X$ denote that $X$ is distributed according to the distribution $\DC_X$. Let the notation $\mathcal{O}$ hide the constant and $\tilde{\mathcal{O}}$ hide constant and polylogarithmic factors of the number of iterations $T$, respectively.
Let $A \oplus B = \{a + b \vert a \in A, b \in B\}$ denote the Minkowski sum of two sets of position vectors $A$ and $B$ in Euclidean space. 

\section{Problem formulation}\label{sec:preliminaries}
Consider the cost function $J(x,\xi) : \XC \times \Xi 
\rightarrow \mathbb{R} $, where $\xi \subseteq  \Xi $ denotes random noise and $x\in \mathcal{X}$ denotes the decision variable with $\XC \subseteq \mathbb{R}^d$ being the admissible set.  Without loss of generality, we assume that $\XC$ contains the ball of radius $r$ centered at the origin, which is denoted as $r \mathbb{B} \subseteq \mathbb{R}^d $. Denote the diameter of the admissible set $\XC$ as $D_x = \sup_{x,y \in \XC} \| x-y\|$.

\subsection{CVaR} 
We use CVaR as risk measure. Suppose $J(x,\xi)$ has the cumulative distribution function $F(y) = P(J(x,\xi) \le y)$, and is bounded by $U>0$, i.e., $|J(x,\xi)|\le U$. Given a confidence level $\alpha \in (0,1]$, the $\alpha$-VaR is
\begin{equation*}
    J^\alpha = F^{-1}(\alpha):= \inf \{y: F(y) \geq \alpha \}.
\end{equation*}
The $\alpha$-CVaR describes the expectation of the $\alpha$-fraction of the worst outcomes of $J(x,\xi)$, and is defined as
\begin{align*}
   C(x) & :=\mathrm{CVaR}_{\alpha}\left[J(x,\xi)\right] \nonumber \\
 & =  \mathbb{E}_F\left[J(x,\xi) \vert J(x,\xi) \geq J^{\alpha}\right]. 
\end{align*}

\subsection{Non-stationary distribution}
Non-stationary stochastic optimization requires some measure to model temporal uncertainties of the dynamic environment. A classic metric is the sum of the variations of the cost functions over time.
In practice, environmental fluctuations might be more intuitively modeled as the time-varying distribution.
Hence, we use the Wasserstein distance metric to quantify distribution variations.  

Wasserstein distance is a distance function that describes the dissimilarity between probability distributions on a certain metric space, see \cite{kantorovich1960mathematical,edwards2011kantorovich}. Let $(\Omega,d)$ be a probability space, where $\Omega$ is a set and $d$ is a metric on $\Omega$. Let $\DC_x$ and $\DC_y$ be two probability distribution on $\Omega$, the dual form of the Wasserstein distance
is given as follows.  
 
\begin{lemma}\cite{edwards2011kantorovich}
    For any fixed $K >0$, 
\begin{equation*}
    W_1(\DC_x, \DC_y)=\frac{1}{K} \sup _{\|f\|_L \leq K} \{ \mathbb{E}_{x \sim \DC_x}[f(x)]-\mathbb{E}_{y \sim \DC_y}[f(y)] \},
\end{equation*}
where $\|\cdot\|_L$ is the Lipschitz norm, 
The right-hand side is the Kantorovich--Rubenstein dual form of the Wasserstein distance metric.
\end{lemma}
As mentioned in Section~\ref{sec:introduction}, some works quantify regret achieved by algorithm using the function variation, i.e., $\sum_{t=2}^T \sup_{x \in \XC}|f_t(x) - f_{t-1}(x)| $, which measures the temporal changes of the function values over time. Inspired by this definition, we introduce the concept of distribution variation as follows.
\begin{definition}\textbf{(Distribution variation)} Let $\{\DC_t\}_{t=1}^T \in \DC$ be the distribution on metric space $\Omega$, with $\DC$ being the admissible set of distribution sequences. 
The distribution variation along iterations  $\{1,\dots,T\}$ is $    \sum_{t=2}^T W_1(\DC_{t-1}, \DC_t)$. 
\end{definition}

\subsection{Problem statement}
Consider the time-varying random noise $\xi_t \sim \DC_t$ and the corresponding cumulative distribution function $F_t$,  the $\alpha$-CVaR of the function $J(x,\xi_t)$ is written as: 
\begin{align*}\label{eq:time varying cvar}
C_t(x) & := \mathrm{CVaR}_{\alpha}\left[J (x,\xi_t )\right] \nonumber \\
& = \mathbb{E}_{F_t}\left[J (x,\xi_t ) \vert J (x,\xi_t ) \geq J^{\alpha}\right].
\end{align*}
We make the following assumptions on the cost function, which are common in the online learning literature, see \cite{hazan2016introduction,besbes2015non}.
\begin{assumption}\label{assumption:convex}
The cost function $J(x,\xi_t) $ is convex in $x$ for every $\xi_t \in \Xi$. 
\end{assumption}
\begin{assumption}\label{assumption:Lipschitz}
The cost function $J(x,\xi_t) $ is Lipschitz continuous in $x$ for every $\xi_t \in \Xi$. That is, there exists a positive real constant $L_0$ such that, for all $x, y \in \XC$, we have $|J(x,\xi_t)-J(y,\xi_t)| \le L_0\|x-y\|$. 
\end{assumption}
It follows the lemma for the CVaR function.
\begin{lemma}\label{lemma:cvar:convex}\cite{cardoso2019risk}
Given Assumption \ref{assumption:convex}, we have that 
$C_t(x)$ is convex in $x$.
\end{lemma}
\begin{assumption}\label{assumption:distribution variation budget} 
The distribution sequence $\{\DC_t\}_{t=1}^T \in \DC$ satisfies the variation budget $V_D$ over the iteration horizon $T$:  $$   \sum_{t=2}^T W_1(\DC_{t-1}, \DC_t) \leq V_D.$$
\end{assumption} 
\begin{assumption}\label{assumption:sublinear budget}
Assume that the distribution variation budget is sub-linear in the iteration horizon $T$, i.e., $V_D = \mathcal{O}(T^{\beta})$ with $\beta \in [0,1)$. 
\end{assumption}
Assumption~\ref{assumption:sublinear budget} enables us to obtain a no-regret, i.e., the upperbound of the regret is sublinear in $T$,  learning policy in stochastic non-stationary environments. 
We provide an example with the non-stationary distribution satisfying Assumptions~\ref{assumption:distribution variation budget} and \ref{assumption:sublinear budget} as follows.
\begin{example}
\cite{doob1942brownian}
Brownian motion is the random motion of particles suspended in a medium. Assuming that $N$ particles start from the origin at the initial time $t=0$, the density of Brownian particles $\rho$ at point $x$ at time $t$ is given as $    \rho(x,t) = \frac{N}{\sqrt{4\pi D t}}e^{-\frac{x^2}{4Dt}}$ 
with $D$ being the mass diffusivity. It can be observed that the distribution flattens with the increasing $t$  and ultimately becomes uniform when time goes to infinity. 
Then the variation of the distribution of random variables $x$ is sub-linear in iteration horizon $T$. 
\end{example}
We use the dynamic regret to measure the performance of the designed algorithm, which is defined as the cumulative loss under the performed actions against the best actions in hindsight: 
\begin{equation}\label{eq:dynamic regret definition}
    \mathrm{DR}(T)  = \sum_{t=1}^T C_t(\hat{x}_t)-  \sum_{t=1}^T C_t(x_t^\ast),
\end{equation}where the action $\hat{x}_t$ is selected according to the designed algorithm at iteration $t$, and $ x_t^\ast = \argmin_{x_t \in \mathcal{X}}C_t (x_t)$ denotes the one-step optimal action at iteration $t$, for $t=1,\dots, T$.  
Specifically, this paper aims to design a risk-averse learning algorithm such that the dynamic regret of this algorithm is bounded in terms of the distribution variation, i.e., $\lim_{T\rightarrow \infty}
\frac{{\rm DR}(T)}{T} = 0 $.

\section{Main Result}\label{sec:Main result}
In this section, we design a risk-averse learning algorithm for both convex and strongly convex cost functions. Then we analyze the dynamic regret using the distribution variation metric.

Before presenting the algorithm, we provide some results for the zeroth-order optimization that lay a foundation for the estimation of CVaR gradient. 
Since the exact CVaR gradient is generally unavailable, we use the zeroth-order optimization algorithm to estimate the CVaR gradient. To begin with, we construct a smoothed approximation of the CVaR function. Given a point $x\in \mathcal{X}$, define the perturbed action by 
$\hat{x}=x+\delta u$, where $u$ is the direction vector sampled from a unit sphere $\mathbb{S}^d \in \mathbb{R}^d$ and $\delta$ is the perturbation radius, also known as the smoothing parameter.  Then, the smoothed version of the CVaR function is given as 
\begin{equation}\label{eq:smoothed cvar}
    C_t^\delta(x) = \mathbb{E}_{u \sim \mathbb{S}^d}[C_t(x+\delta u)].
\end{equation} 
In the following, we present lemmas regarding properties of smoothed approximation of the CVaR function. 
\begin{lemma}\label{lemma:smoothed cvar convexity} \cite{cardoso2019risk} Given Assumption \ref{assumption:convex}, we have that  $C_t^\delta\left(x\right)$ is convex in $x$. 
\end{lemma}
\begin{lemma}\label{lemma:smoothed cvar Lipschitz} \cite{cardoso2019risk} Given Assumption \ref{assumption:Lipschitz}, we have that  $C_t^\delta (x)$ is $L_0$-Lipschitz in $x$ and $|C_t^\delta(x) - C_t(x)| \le \delta L_0 $.
\end{lemma} 
Non-stationary environments require decision makers to continuously adapt to changing conditions. The generic idea of the \textit{restarting procedure} is to let the learning algorithm reset its internal state and update its parameters, and therefore capture the new dynamics in the environment. 
Let $\mathcal{A}$ be an online optimization algorithm. We employ the restarting procedure to refresh the parameters and restart $\mathcal{A}$ every $\Delta_T$ iterations. 

The risk-averse learning algorithm is given as Algorithm~\ref{alg:algorithm}. Suppose that there are totally $T$ iterations and they are divided into $s$ batches with a length of $\Delta_T \in (1,T)$, where $s = \left\lceil\frac{T}{\Delta_T}\right\rceil$. Each batch is formally defined as
\begin{align*}
    \TC_j  =  \big\{ t : (j-1)\Delta_T < t \le \min\{T, j\Delta_T\} \big\},& \nonumber  \\ 
        {\rm for}~j = 1, \dots, s&.
\end{align*} 
For each iteration $t$, inside batch $j$, it holds that $j=\left\lceil\frac{t}{\Delta_T}\right\rceil$ and its timestamp within the batch is epoch $\tau$, i.e.,
\begin{align}\label{eq:epoch}
    \tau = t-(j-1) \Delta_T.
\end{align}
\begin{algorithm}[t] 
\caption{Risk-averse learning with restarting procedure} \label{alg:algorithm}
\begin{algorithmic}[1]
    \REQUIRE Initial value $x_0$, iteration horizon $T$, batch size $\Delta_T$, smoothing parameter $\delta$,  risk level $\alpha$.
    \FOR{$ {\rm{iteration}} \;t = 1,\dots, T$} 
    \STATE Identify batch $j = \left\lceil \frac{t}{\Delta_T} \right \rceil$
    \STATE Identify epoch in batch $ \tau = t -  (j-1)\Delta_T $
    \STATE Select sampling number $n_t = \phi(\tau)$ and learning rate $\eta_t = \sigma(\tau)$
    \STATE  Sample $u_{t} \in \mathbb{S}^{d}$
    \STATE  Play $\hat{x}_{t}=x_{t}+\delta u_{t} $
            \FOR{$i=1,\ldots,n_t$}
            \STATE Play $\hat{x}_{t}$ and obtain $J_t(\hat{x}_{t},\xi_t^i)$
            \ENDFOR
            \STATE Build empirical distribution function $\hat{F}_{t}(y)$ given in \eqref{eq:EDF}
            \STATE Estimate CVaR: $ {\rm{CVaR}}_{\alpha}[\hat{F}_{t}] $ 
            \STATE Construct gradient estimate\\ $\hat{g}_{t}=\frac{d}{\delta}   {\rm{CVaR}}_{\alpha} [\hat{F}_{t} ] u_{t}$
            \STATE Update $x$: $x_{t+1} \leftarrow \mathcal{P}_{\mathcal{X}_t^{\delta}} ( x_{t} - \eta_{\tau} \hat{g}_{t})$
    \ENDFOR
\end{algorithmic}
\end{algorithm}Then, we design the sampling strategy depending on the epoch $\tau$, which is given as $n_t = \phi(\tau)$. We allow the algorithm to sample the cost function values multiple times at each iteration, therefore improving the estimation accuracy under the changing distribution. The sampling strategy function $\phi $ satisfies 
\begin{align}\label{eq:sampling requirement}
    \sum_{\tau=1}^{\Delta_T} \frac{1}{\sqrt{\phi(\tau)}} \le c \Delta_T^{1-\frac{a}{2}}
\end{align}
with the tuning parameter $a>0$ and some constant $c>0$. 
The sampling strategy proposed in \cite{wang2022zeroth} is an example that satisfies \eqref{eq:sampling requirement},
where 
\begin{equation}\label{eq:sampling example}
    \phi(\tau) = \left\lceil b (\Delta_T-\tau+1)^a\right\rceil
\end{equation}
with parameters $a,b>0$. Specifically, the tuning parameter $a$ plays the same role in \eqref{eq:sampling requirement} and \eqref{eq:sampling example}, where a higher value of $a$ corresponds to a larger number of sampling numbers over the iteration horizon.
Similarly, the learning rate at each batch is designed according to $\eta_t = \sigma(\tau)$, 
where the function $\sigma$ will be designed later. In Algorithm~\ref{alg:algorithm}, we refresh the evolution of the sampling number $n_t$ and the learning rate $\eta_t$ every $\Delta_T$ iterations.


At iteration $t$, we implement the perturbed action $\hat{x}_t = x_t +\delta u_t$ for $n_t$ times and obtain the cost function values denoted by $J(\hat{x}_t,\xi^i),~i=1,\ldots,n_t$. 
Then, we use the queried function values to construct the empirical distribution function, which is given as
\begin{equation}\label{eq:EDF}
    \hat{F}_t(y)=\frac{1}{n_t} \sum_{i=1}^{n_t} \mathbf{1} \{J(\hat{x}_t, \xi_t^i) \leq y \}.
\end{equation} 
With this empirical distribution function, we construct the CVaR estimate $\cvar_\alpha[\hat{F}_t]$ and further the CVaR gradient estimate 
\begin{equation}\label{eq:estimated gradient}
    \hat{g}_ t=\frac{d}{\delta} \cvar_\alpha\big[\hat{F}_t\big] u_t.
\end{equation}
The gradient descent update for the risk-averse learning process proceeds as follows: 
\begin{equation}\label{eq:gradient descent}
    x_{t+1} = \PC_{\XC^\delta}(x_t - \eta_t \hat{g}_t),
\end{equation} 
where $\PC_{\XC^\delta }(x):= \argmin_{y\in \XC^\delta}\|x - y\|^2$ denotes the projection operator with $\XC^\delta = \{x \in \XC \vert \frac{1}{1-\delta/r} x \in \XC\}$ being the  projection set. The projection keeps the sampled actions $\hat{x}_t$ inside of the admissible set $\XC$, which establishes as
\begin{align*}
   \bigg(1-\frac{\delta}{r}\bigg)\XC \oplus \delta   \mathbb{B} &= \bigg(1-\frac{\delta}{r}\bigg)\XC \oplus \frac{\delta}{r} r   \mathbb{B}\\
 & \subseteq \bigg(1-\frac{\delta}{r}\bigg)\XC \oplus \frac{\delta}{r} \XC = \XC.  
\end{align*}
Without loss of generality, we let the initial action at the restarting epoch be based on action learned from the previous batch.  

Note that we construct the empirical distribution function of the cost function using finite samples, which induces the CVaR gradient estimate error: 
\begin{equation}\label{eq:CVaR estimation error}
    \hat{\epsilon}_t:= \cvar_\alpha[\hat{F}_t] - \cvar_\alpha[F_t].
\end{equation}
In the following, we present two lemmas to bound the estimate error. The first lemma shows that the CVaR values with two cumulative distribution functions can be bounded by the sup difference of these two cumulative distribution functions, which is presented below.
\begin{lemma}\label{lemma:cvar-estimation error bound} \cite{wang2022zeroth} Let $F$ and $G$ be two cumulative distribution functions of two random variables and the random variables are bounded by $U$. Then we have that \begin{equation}\label{eq:cvar bound}
        |\cvar_\alpha[F] -\cvar_\alpha [G] | \le \frac{U}{\alpha} \sup_{x} |F(x)-G(x)|.
    \end{equation}
\end{lemma}
The following lemma shows the fluctuation of CVaR values is bounded in terms of the distribution shift.
\begin{lemma}\label{lemma:cva-wasserstein}
    Suppose $f(x)$ is $L_0$-Lipschitz in $x$. For two random variables $X$ and $Y$ with distributions $\DC_X$ and $\DC_Y$, respectively, we have
\begin{equation}\label{eq:cvar-wasserstein}
    |\mathop{\cvar_\alpha} \limits_{X \sim \DC_X } [f(X)]- \mathop{\cvar_\alpha} \limits_{Y \sim \DC_Y } [f(Y)]\vert \le \frac{L_0}{\alpha} W_1(\DC_X,\DC_Y).
\end{equation}
\end{lemma}
The proof of Lemma~\ref{lemma:cva-wasserstein} is provided in the Appendix.

\subsection{Convex case} 
In this section, we investigate the dynamic regret of Algorithm~\ref{alg:algorithm} for the convex case. The main result is presented in the following theorem.
\begin{theorem}\label{theorem:convex}
Let Assumptions \ref{assumption:convex}, \ref{assumption:Lipschitz}, and \ref{assumption:distribution variation budget} hold.  Suppose that the sampling numbers over iteration horizon $T$ satisfy \eqref{eq:sampling requirement} with a constant $a>0$. 
\begin{enumerate}
    \item When $a \in (0,1]$, select $  \delta = \left(\frac{V_D}{T}\right)^{\frac{a}{4+a}}$, $  \eta_t = \left(\frac{V_D}{T}\right)^{\frac{3a}{4+a}}$, $\Delta_T = \left(\frac{T}{V_D}\right)^{\frac{4}{4+a}} $. Then,   
 Algorithm~\ref{alg:algorithm}  achieves ${\rm DR}(T) =  \tilde{\mathcal{O}}(T^{\frac{4}{4+a}}V_D^{\frac{a}{4+a}}) $ with high probability.
 \item When $a >1$, select  $  \delta = \left(\frac{V_D}{T}\right)^{\frac{1}{5}}$, $  \eta_t = \left(\frac{V_D}{T}\right)^{\frac{3}{5}}$, $\Delta_T = \left(\frac{T}{V_D}\right)^{\frac{4}{5}} $. Then, Algorithm~\ref{alg:algorithm}  achieves ${\rm DR}(T) =  \tilde{\mathcal{O}}(T^{\frac{4}{5}}V_D^{\frac{1}{5}}) $ with high probability.
\end{enumerate}
\end{theorem}
\textit{Proof.}
For $t = 1,\dots,T$, we have 
\begin{align}\label{eq:projection mismatch}
   \min_{x_t \in \mathcal{X}^\delta} C_t^\delta(x_t) &=    \min_{x_t \in \mathcal{X}} C_t^\delta((1-\delta/r)x_t) \nonumber \\
   &\leq  \min_{x_t \in \mathcal{X}} (\delta/r) C_t^\delta(0) + (1-\delta/r)C_t^\delta(x_t) \nonumber \\
    &\leq \min_{x_t\in \mathcal{X}} C_t^\delta(x_t)+ (\delta/r)L_0 \left\|x_t \right\| \nonumber \\
   &\leq \min_{x_t\in \mathcal{X}} C_t^\delta(x_t)+D_x L_0 \delta/r   .
\end{align}
The first inequality is from the convexity of $C_t^\delta(x)$ as shown in Lemma~\ref{lemma:smoothed cvar convexity}, and the second inequality is from Lipschitzness of $C_t^\delta(x)$ as shown in Lemma \ref{lemma:smoothed cvar Lipschitz}.
To simplify notations, we denote $x_t^{\delta,\ast} = \argmin_{x \in \mathcal{X}^\delta} C_t^\delta (x)$. 
The dynamic regret defined in \eqref{eq:dynamic regret definition} is further written as 
\begin{align}\label{eq:convex1}
        \mathrm{DR}(T)  &\leq \sum_{t=1}^T C_t^\delta( \hat{x}_t)- \sum_{t=1}^T C_t^\delta (x_t^\ast)+2 \delta L_0 T \nonumber \\ 
        &\leq \sum_{t=1}^T C_t^\delta( x_t)- \sum_{t=1}^T C_t^\delta (x_t^\ast)+3 \delta L_0 T \nonumber \\ 
    &\leq \sum_{t=1}^T  C_t^\delta(x_t)-  \sum_{t=1}^T C_t^\delta (x_t^{\delta,\ast})+(3 +D_x/r)\delta L_0 T ,\nonumber \\ 
\end{align}
where the first inequality follows from the definition of the function $C_t^\delta$ as in \eqref{eq:smoothed cvar}, the second inequality follows from the Lipschitzness of the function $C_t^\delta$, and the third inequality establishes by substituting \eqref{eq:projection mismatch} into the $ C_t^\delta (x_t^\ast)$. Denote
$x_j^{\delta,\ast} = \argmin_{x \in \mathcal{X}^\delta} \sum_{t\in \TC_j} C_t^\delta (x)$ as the single best action over batch $j$, for $j = 1, \dots, s$. 
It follows that 
\begin{align}\label{eq:DR fraction}
&\hspace{1em} \sum_{t=1}^T  C_t^\delta(x_t)-  \sum_{t=1}^T C_t^\delta (x_t^{\delta,\ast})  \nonumber \\ 
 &=\sum_{j=1}^s \sum\limits_{t \in \TC_j} \Big( C_t^\delta(x_t)-  C_t^{\delta}(x_j^{\delta,\ast}) \Big) \nonumber \\
& \hspace{1em}+ \sum_{j=1}^s \sum\limits_{t \in \TC_j}  \Big(  C_t^{\delta}(x_j^{\delta,\ast}) - C_t^\delta (x_t^{\delta,\ast}) \Big)\nonumber \\ 
&= \sum_{j=1}^s \RC_1^j + \RC_2,
\end{align} 
with $  \RC_1^j= \sum\limits_{t \in \TC_j} \Big(C_t^\delta(x_t)-  C_t^{\delta}(x_j^{\delta,\ast}) \Big) $, for $j=1,\dots,s$,   and $\RC_2=\sum_{j=1}^s \sum\limits_{t \in \TC_j}  \Big(  C_t^{\delta}(x_j^{\delta,\ast}) - C_t^\delta (x_t^{\delta,\ast})\Big)$. 
We first bound the term $\RC_1^j $.
By the convexity of the function $C_t^\delta$, we have  
\begin{equation*}
     \RC_1^j   \le \sum_{t\in \TC_j} \langle \nabla C_t^\delta(x_t) ,x_t - x_j^{\delta,\ast} \rangle.
\end{equation*}
In combination of \eqref{eq:estimated gradient} and \eqref{eq:CVaR estimation error}, we obtain that 
\begin{equation}\label{eq:cvar derivative}
   \nabla C_t^\delta(x_t)   = \mathbb{E}[\hat{g}_t - \frac{d}{\delta}\hat{\epsilon}_tu_t].
\end{equation}
By the update rule \eqref{eq:gradient descent}, we have 
\begin{align}
  &  \hspace{1.3em}\|x_{t+1} - x_j^{\delta,\ast} \|^2 \nonumber\\
  &= \|\PC_{\XC^\delta }(x_{t}-\eta_t \hat{g}_t) - x_j^{\delta,\ast} \|^2 \nonumber \\
 &\leq  \| x_{t}-\eta_t \hat{g}_t  - x_j^{\delta,\ast}\|^2 \nonumber \\ 
& = \|x_t-x_j^{\delta,\ast}\|^2+\eta_t^2\|\hat{g}_t\|^2-2\eta_t \langle \hat{g}_t, x_t-x_j^{\delta,\ast} \rangle  \nonumber ,
\end{align}
where the inequality follows from the fact that $x_j^{\delta,\ast} \in \mathcal{X}^{\delta}$. 
Then, we obtain
\begin{align}\label{eq:gt-x}
  &\hspace{1em} \langle \hat{g}_t, x_t-x_j^{\delta,\ast} \rangle \nonumber\\
   &\leq \frac{\|x_t-x_j^{\delta,\ast}\|^2 -\|x_{t+1}-x_j^{\delta,\ast}\|^2}{2\eta_t} + \frac{\eta_t}{2}\|\hat{g}_t\|^2.
\end{align}
Substituting \eqref{eq:cvar derivative} and \eqref{eq:gt-x} into $\RC_1^j$, we
obtain 
\begin{align}\label{eq:R1j}
\RC_1^j  &\leq  \sum_{t\in \TC_j} \mathbb{E}[\langle \hat{g}_t - \frac{d}{\delta}\hat{\epsilon}_t, x_t - x_j^{\delta,\ast} \rangle]    \nonumber \\
 &\leq \sum_{t\in \TC_j} \Big( \frac{1}{2\eta_t} \mathbb{E}[ \|x_t-x_j^{\delta,\ast}\|^2 -\|x_{t+1}-x_j^{\delta,\ast}\|^2  ]   \nonumber \\
&\hspace{1em}+ \frac{\eta_t}{2}\mathbb{E}[\|\hat{g}_t\|^2]  +\frac{d}{\delta}\mathbb{E}[\|\hat{\epsilon}_t\|\| x_t -x_j^{\delta,\ast}\| ] \Big)\nonumber  \\
&= \frac{1}{2\eta_t}\left( 
 \| x_{(j-1)\Delta_T+1}  -  x_j^{\delta,\ast}\|^2 -  \|x_{j\Delta_T} - x_j^{\delta,\ast}\|^2 \right) \nonumber \\
 &\hspace{1em}+ \RC_{12}^j + \RC_{13}^j \nonumber \\
 &\leq \frac{D_x^2}{\eta_t}  + \RC_{12}^j + \RC_{13}^j,
\end{align}
with $ \RC_{12}^j =\sum\limits_{t\in \TC_j} \frac{\eta_t}{2}\mathbb{E}[\|\hat{g}_t\|^2] $ and $ \RC_{13}^j =  \sum\limits_{t\in \TC_j} \frac{d}{\delta} \mathbb{E}[\|\hat{\epsilon}_t\|\| x_t -x_j^{\delta,\ast}\| ] $.  
Regarding $ \RC_{12}^j$, for $i = 1,\dots, s$, it writes, 
\begin{align}\label{eq:R12j}
 \RC_{12}^j &=  \sum_{t\in \TC_j} \frac{\eta_t}{2} \left\|\frac{d}{\delta} \mathrm{CVaR}_{\alpha}[\hat{F}_{ t}] u_{t} \right\|^2 \nonumber \\
 &\leq  \sum_{t\in \TC_j} \frac{\eta_t}{2}  \left(\frac{d U}{\delta}\right)^2  \le  \frac{d^2 U^2 \eta_t }{2\delta^2}   \Delta_T   .
\end{align} 
The first inequality establishes as $\cvar_\alpha[\hat{F}_{ t}] \le U$.  Regarding $ \RC_{13}^j$, we first analyze  the bound of $\hat{\epsilon}_t$ given in \eqref{eq:CVaR estimation error}. By leveraging the Dvoretzky–Kiefer–Wolfowitz (DKW) inequality \cite{dvoretzky1956asymptotic}, we have that 
\begin{equation}\label{eq:DKW}
    \mathbb{P}\left\{ \sup_{y} |\hat{F}_t(y) - F_t(y)| \ge \sqrt{\frac{\ln (2 / \bar{\gamma})}{2 n_t}}  \right\} \le  \bar{\gamma}.
\end{equation}
Denote the event in \eqref{eq:DKW} as $A_t$, and $\mathbb{P}\{A_t\}$ denotes the occurrence probability of event $A_t$, for $t=1,\dots, T$.
By Lemma~\ref{lemma:cvar-estimation error bound}, we bound the error of CVaR estimate by 
\begin{align}
\label{eq:DKW cvar mismatch}
     |\hat{\varepsilon}_{t} |  = \frac{U}{\alpha}\sup\big| \hat{F}_t - F_t\big|  \le \frac{U}{\alpha} \sqrt{\frac{\ln (2 / \bar{\gamma})}{2 n_t}} 
\end{align}
with probability at least $1 -\bar{\gamma}$, for $t = 1,\dots, T$.  Let $\gamma = \bar{\gamma}T$. 
Then, by substituting \eqref{eq:DKW cvar mismatch} into $\RC_{13}^j$, we obtain that  
\begin{align}\label{eq:R13j}
\sum_{j=1}^s\RC_{13}^j  &\leq  \frac{d D_x}{ \delta}\sum_{j=1}^s \sum_{t \in \TC_j} \mathbb{E}[\|\hat{\varepsilon}_t\|]  \nonumber \\
 &\leq    \frac{d U D_x}{\alpha \delta}     \sum_{j=1}^s\sum_{t \in \TC_j} \sqrt{\frac{  \ln (2 T / \gamma)}{2n_t}}    \nonumber \\ 
 &\leq \frac{cd U D_x}{\alpha \delta} \left\lceil \frac{T}{\Delta_T}\right\rceil \sqrt{\frac{  \ln (2 T / \gamma)}{2}} \Delta_T^{1-\frac{a}{2}}  
\end{align} 
with probability at least $1-\gamma$, which establishes as $    1 - \mathbb{P}\{\bigcup_{t=1}^{T} A_t \} \ge 1 - \sum_{t=1}^{T} \mathbb{P}\{A_t\} \ge 1 - T\frac{\gamma}{T} \ge 1-\gamma$. As shown in \eqref{eq:DKW cvar mismatch}, when the number of samples increases, the empirical density function approaches the true one. Namely, a sufficient number of samples enable the event $\{\bigcup_{t=1}^{T} A_t\}$ to occur with high probability.
Regarding $\RC_2$, we have that 
\begin{align}\label{eq:R2 distribution}
        \RC_2  
     &\leq \frac{2L_0}{\alpha} \Delta_T  \sum_{t=2}^T W_1(\DC_t,\DC_{t-1}) \leq \frac{2L_0}{\alpha}\Delta_TV_D,
\end{align}
where the first inequality follows from Lemma~\ref{lemma:function variation}, respectively.
The second inequality follows from Lemma~\ref{lemma:cva-wasserstein}, and the last inequality follows from Assumption~\ref{assumption:distribution variation budget}. 
Moreover, Lemma~\ref{lemma:cva-wasserstein} indicates that when batch size $\Delta_T$ approaches $1$, the batch-optimal actions $x_j^{\delta,\ast}$ will be closer to one-step optimal action $x_t^{\delta,\ast}$, for $t \in \TC_j$, and the accumulated loss defined by \eqref{eq:R2 distribution} will be smaller. However, restarting also induces errors such as $ \frac{D_x^2}{\eta_t} $. Thus, it is necessary to select an optimal batch size  $\Delta_T$  to minimize the regret. 
Substituting \eqref{eq:R1j}, \eqref{eq:R12j}, \eqref{eq:R13j} and \eqref{eq:R2 distribution} into \eqref{eq:DR fraction}, and combining it with \eqref{eq:convex1},  it results   
\begin{align}\label{eq:regret minimization convex}
   & {\rm DR}(T) \nonumber \\ 
     &\leq  (3 +D_x/r)\delta L_0 T +  \RC_2  +\sum_{j=1}^s\left(\frac{D_x^2}{\eta_t}  + \RC_{12}^j + \RC_{13}^j \right) \nonumber \\ 
    &\leq  (3 +D_x/r)\delta L_0 T + \frac{2L_0}{\alpha}\Delta_T V_D  +\frac{D_x^2}{\eta_t}  \left\lceil \frac{T}{\Delta_T} \right\rceil \nonumber \\
    &    + \left(     \frac{d^2 U^2 \eta_t \Delta_T}{2\delta^2}    + \frac{cd U D_x }{\alpha \delta} \sqrt{\frac{  \ln (2 T / \gamma)}{2}}  \Delta_T^{1-\frac{a}{2}}   \right) \left\lceil \frac{T}{\Delta_T} \right\rceil   
    \nonumber \\ 
     &\leq (3 +D_x/r)\delta L_0 T + \frac{2L_0}{\alpha}\Delta_T V_D  +\frac{2D_x^2T}{\eta_t\Delta_T}   \nonumber \\
    &    +      \frac{d^2 U^2 \eta_t  T}{\delta^2}    + \frac{\sqrt{2}cd U D_x \sqrt{  \ln (2  T / \gamma) }}{\alpha \delta}   T\Delta_T^{-\frac{a}{2}}.      
\end{align}
with probability at least $ 1- \gamma $.
The last inequality results from $\left\lceil \frac{T}{\Delta_T} \right\rceil   \le\frac{T}{\Delta_T} +1 \le \frac{2T}{\Delta_T} $. 
When $a \in (0,1]$, we select $  \delta = \left(\frac{V_D}{T}\right)^{\frac{a}{4+a}}$, $  \eta_t = \left(\frac{V_D}{T}\right)^{\frac{3a}{4+a}}$ and $\Delta_T = \left(\frac{T}{V_D}\right)^{\frac{4}{4+a}} $ to minimize the dynamic regret, it achieves  ${\rm DR}(T) =  \tilde{\mathcal{O}}(T^{\frac{4}{4+a}}V_D^{\frac{a}{4+a}}) $ with probability at least $ 1- \gamma $ (see \cite{flaxman2004online} for parameters selection details).      When $a >1$, we select  $  \delta = \left(\frac{V_D}{T}\right)^{\frac{1}{5}}$, $  \eta_t = \left(\frac{V_D}{T}\right)^{\frac{3}{5}}$ and $\Delta_T = \left(\frac{T}{V_D}\right)^{\frac{4}{5}} $, it achieves ${\rm DR}(T) =  \tilde{\mathcal{O}}(T^{\frac{4}{5}}V_D^{\frac{1}{5}}) $ with probability at least $ 1- \gamma $. The proof is complete.  \hfill $\qed$
\begin{remark} 
When $a\in(0,1]$, the regret bound achieved by Algorithm~\ref{alg:algorithm} decreases with the increasing tuning parameter $a$ and ceases increasing when $a>1$. It is because the dynamic regret contains the accumulative error of CVaR gradient estimates, i.e., $ \frac{d^2 U^2 \eta_t  T}{\delta^2}    + \frac{\sqrt{2}cd U D_x \sqrt{  \ln (2  T / \gamma) }}{\alpha \delta}   T\Delta_T^{-\frac{a}{2}}$, which is induced by using finite samples to estimate the CVaR gradients in the zeroth-order algorithm. Since a larger tuning parameter $a$ implies more queries of cost values and a more accurate estimate of CVaR gradients, the order of the accumulative estimation error decreases with the increasing $a$. Furthermore, we select the smoothing parameter $\delta$, the batch size $\Delta_T$ and the step size $\eta_t$ that minimizes the regret defined in \eqref{eq:regret minimization convex}. When $a\in(0, 1]$, this accumulative estimation error term is one of the dominant terms in  \eqref{eq:regret minimization convex}. While when $a > 1$, it becomes negligible compared with the rest terms in \eqref{eq:regret minimization convex}. 
\end{remark}
\subsection{Strongly convex case}
In the section, we further investigate Algorithm~\ref{alg:algorithm} for the strongly convex case. We first provide the following assumption and lemma related to the strongly convex condition, which are common in risk-averse learning, see \cite{anderson2020varying,cardoso2019risk} 
\begin{assumption}\label{assumption:strongly convex}
The function $J(x,\xi_t): \XC \times \Xi \rightarrow \mathbb{R} $ is $m$-strongly convex in $x$ for every $\xi_t \in \Xi$. That is,  for all $x, y \in \XC $ and every $\xi_t \in \Xi$, we have 
$$J(y,\xi_t) \ge J(x,\xi_t) + \nabla_x J(x,\xi_t)^\top(y-x) + \frac{m}{2}\| x-y\|_2^2.$$
\end{assumption}
It follows the lemma for the CVaR function and the smoothed version of the CVaR function. 
\begin{lemma}\label{lemma:strongconvex}
Given Assumption \ref{assumption:strongly convex}, we have that 
\begin{enumerate}
    \item $C_t(x)$ is $m$-strongly convex in $x$;
    \item $C_t^{\delta}(x)$ is $m$-strongly convex in $x$.
\end{enumerate}
\end{lemma} 
The proof of Lemma~\ref{lemma:strongconvex} is provided in the Appendix. 
Now we are ready to present the main result for the strongly convex case.
\begin{theorem}\label{theorem:strongly convex}
Let Assumptions  \ref{assumption:Lipschitz}, \ref{assumption:distribution variation budget}  and \ref{assumption:strongly convex} hold. 
Suppose that the sampling numbers over iteration horizon $T$ satisfy  \eqref{eq:sampling requirement} with a constant $a>0$.
Select the learning rate as 
\begin{equation}\label{eq:eta_t}
\eta_t = \sigma(\tau) = \frac{1}{m\tau},
\end{equation}
where the epoch $\tau$ is obtained from \eqref{eq:epoch}.
\begin{enumerate}
    \item When $a \in (0,\frac{4}{3}]$,  select $  \delta = \left(\frac{V_D}{T}\right)^{\frac{a}{4+a}}$ and $\Delta_T = \left(\frac{T}{V_D}\right)^{\frac{4}{4+a}} $. Then, 
 Algorithm~\ref{alg:algorithm} achieves dynamic regret ${\rm DR}(T) =  \tilde{\mathcal{O}}(T^{\frac{4}{4+a}}V_D^{\frac{a}{4+a}}) $ with high probability; 
 \item When $a > \frac{4}{3} $, select $  \delta = \left(\frac{V_D}{T}\right)^{\frac{1}{4}}$ and $\Delta_T = \left(\frac{T}{V_D}\right)^{\frac{3}{4}} $. Then Algorithm~\ref{alg:algorithm} achieves dynamic regret ${\rm DR}(T) =  \tilde{\mathcal{O}}(T^{\frac{3}{4}}V_D^{\frac{1}{4}}) $ with high probability.
 \end{enumerate}
\end{theorem}
\textit{Proof.}
Following the derivation of \eqref{eq:projection mismatch}-\eqref{eq:DR fraction} in the proof of Theorem~\ref{theorem:convex}, the dynamic regret under Algorithm~\ref{alg:algorithm} is written as 
\begin{align}\label{eq:DR1 strongly}
  {\rm DR}(T)  
&\leq (3 +D_x/r)\delta L_0 T +   \sum_{j=1}^s \sum\limits_{t \in \TC_j} \left(C_t^\delta(x_t)-  C_t^{\delta}(x_j^{\delta,\ast}) \right) \nonumber \\ & \hspace{1em}+   \sum_{j=1}^s \sum\limits_{t \in \TC_j}    \left(C_t^{\delta}(x_j^{\delta,\ast}) - C_t^\delta (x_t^{\delta,\ast})\right) \nonumber \\
&\leq (3 +D_x/r)\delta L_0 T + \sum_{j=1}^s\tilde{\RC}_1^j + \RC_2
\end{align} 
with $\tilde{\RC}_1^j = \sum\limits_{t \in \TC_j} C_t^\delta(x_t)-  C_t^{\delta}(x_j^{\delta,\ast})$ and the definition of $\RC_2$ is as in \eqref{eq:DR fraction}.  
By the strong convexity of the function $C_t^\delta$, we have 
\begin{align}\label{eq:R1 strongly}
    \tilde{\RC}_1  &\leq \sum_{j=1}^s \Big( \sum_{t \in \TC_j} \langle \nabla C_t^\delta(x_t), x_t - x_j^{\delta,\ast} \rangle - \frac{m}{2}\|x_t-x_j^{\delta,\ast}\|^2   \Big) \nonumber \\ 
   &\leq \sum_{j=1}^s \sum_{t \in \TC_j} \Big( \frac{1}{2\eta_t}\big(\|x_t - x_j^{\delta,\ast}\|^2 - \|x_{t+1}-  x_j^{\delta,\ast}\|^2)  \nonumber \\
 & \hspace{1em} - \frac{m}{2}\|x_t-x_j^{\delta,\ast}\|^2 + \frac{\eta_t}{2}\mathbb{E}[\|\hat{g}_t\|^2] \nonumber \\
 &\hspace{1em}+ \frac{d}{\delta}\mathbb{E}[\|\hat{\epsilon}_t\|\| x_t -x_j^{\delta,\ast} \|] \Big)\nonumber \\
&\leq \sum_{j=1}^s \Big( \tilde{\RC}_{11}^j  + \tilde{\RC}_{12}^j + \RC_{13}^j \Big)
\end{align} 
where $\tilde{\RC}_{11}^j =  \sum_{t \in \TC_j} \Big( \frac{1}{2\eta_t}\big(\|x_t - x_j^{\delta,\ast}\|^2 - \|x_{t+1}-  x_j^{\delta,\ast}\|^2 \big)  - \frac{m}{2}\|x_t-x_j^{\delta,\ast}\|^2 \Big)$, $ \Tilde{\RC}_{12}^j = \sum_{t\in \TC_j} \frac{\eta_t}{2}\mathbb{E}[\|\hat{g}_t\|^2] $, and the definition of $\RC_{13}^j$ is the same as \eqref{eq:R13j}. The second inequality follows the derivation of \eqref{eq:cvar derivative}-\eqref{eq:R1j} in Theorem~\ref{theorem:convex}. For notational simplicity, denote the first epoch of $\TC_j$ as $\bar{\tau}_j$. By expanding the sequences of $\tilde{\RC}_{11}^j $,  we obtain 
\begin{align}\label{eq:R11 strongly}
\sum_{j=1}^s \tilde{\RC}_{11}^j  
&= \sum_{j=1}^s  \sum_{t \in \{\TC_j\backslash \bar{\tau}_j\}} \big(\frac{1}{2\eta_{t}} - \frac{1}{2\eta_{t-1}} - \frac{m}{2} \big)    \|x_t - x_j^{\delta,\ast}\|^2   \nonumber \\
&\hspace{1em}+ \frac{1}{2\eta_{\bar{\tau}_j}}\|x_{\bar{\tau}_j}-x_j^{\delta,\ast} \|^2 -\frac{1}{2\eta_{j\Delta_T}}\|x_{j\Delta_T}-x_j^{\delta,\ast} \|^2 \nonumber \\
&\hspace{1em}- \frac{m}{2}\|x_{\bar{\tau}_j} - x_j^{\delta,\ast}\|^2\nonumber \\
&\leq  \sum_{j=1}^s \frac{1}{2\eta_{\bar{\tau}_j}}\|x_{\bar{\tau}_j}-x_j^{\delta,\ast} \|^2  \le m D_x^2\left\lceil \frac{T}{\Delta_T} \right\rceil.
\end{align}
The first inequality establishes by substituting the learning rate given by \eqref{eq:eta_t} into $\tilde{\RC}_{11}^j$, and $\frac{m}{2}, \frac{1}{2\eta_{j\Delta_T}} > 0$. 
Regarding $\sum_{j=1}^s \Tilde{\RC}_{12}^j$, it  writes
 \begin{align}\label{eq:strongly R12j}
\sum_{j=1}^s \Tilde{\RC}_{12}^j &\leq \frac{d^2 U^2}{2\delta^2} \sum_{j=1}^s  \sum_{t\in \TC_j}  \eta_t  \leq   \frac{d^2 U^2  }{2\delta^2 m} \sum_{j=1}^s \sum_{t=1}^{\Delta_T}\frac{1}{t}   \nonumber \\
&\leq  \frac{d^2 U^2  }{2\delta^2 m}(1+\ln \Delta_T   )\left\lceil \frac{T}{\Delta_T} \right\rceil.
\end{align} 
The last inequality establishes as $\sum_{t=1}^{\Delta_T}\frac{1}{t}  \le 1+\ln \Delta_T$. 
Note that the upperbound of $\RC_2$ only relates to the distribution variation budget, which applies here directly. Hence, we substitute \eqref{eq:R13j}, \eqref{eq:R11 strongly} and \eqref{eq:strongly R12j} into \eqref{eq:R1 strongly}, combine them with \eqref{eq:DR1 strongly}, and obtain 
\begin{align}\label{eq:regret minimization strongly}
    {\rm DR}(T)
    &=  (3 +D_x/r)\delta L_0 T + \RC_2 + \sum_{j=1}^s\tilde{\RC}_{11}^j+  \tilde{\RC}_{12}^j + \RC_{13}^j  \nonumber \\ 
    &\leq (3 +D_x/r)\delta L_0 T + \frac{2L_0}{\alpha}\Delta_TV_D +  m D_x^2  \left\lceil \frac{T}{\Delta_T}\right\rceil  \nonumber \\ 
    & \hspace{1em}+ \frac{d^2 U^2  }{2\delta^2 m}(1+\ln \Delta_T   )\left\lceil \frac{T}{\Delta_T} \right\rceil\nonumber \\ 
    & \hspace{1em} + \frac{cd U D_x}{\alpha \delta} \sqrt{\frac{ \ln (2 T / \gamma)}{2}}  \Delta_T^{1-\frac{a}{2}} \left\lceil \frac{T}{\Delta_T}\right\rceil \nonumber \\ 
    &\leq (3 +D_x/r)\delta L_0 T + \frac{2L_0}{\alpha}\Delta_TV_D + 2 m D_x^2  \frac{T}{\Delta_T}   \nonumber \\ 
    & \hspace{1em} + \frac{d^2 U^2  }{ \delta^2 m}(1+\ln T   )\frac{T}{\Delta_T}  \nonumber \\ 
    &\hspace{1em} + \frac{\sqrt{2}cd U D_x\sqrt{ \ln (2 T / \gamma) }}{\alpha \delta}  T \Delta_T^{ -\frac{a}{2}},  
\end{align}
with probability at least $1-\gamma$
The remaining claim is as in Theorem~\ref{theorem:strongly convex}. The proof completes here.  \hfill $\qed$

\begin{remark}
Comparing the results of Theorems~\ref{theorem:convex} and \ref{theorem:strongly convex}, we observe that Algorithm~\ref{alg:algorithm} achieves the same regret order in the strongly convex and convex cases when $a\in(0,1]$, and achieves smaller regret in the strongly convex case than in the convex case when $a>1$. Additionally, in the strongly convex case, this regret order reduction ceases when $a > \frac{4}{3}$. 
This is because the accumulative error of CVaR gradient estimates in the strongly convex case is $   \frac{d^2 U^2  }{ \delta^2 m}(1+\ln T   )\frac{T}{\Delta_T}  + \frac{\sqrt{2}d U D_x\sqrt{ \ln (2 \Delta_T / \gamma) }}{\alpha \delta}  T \Delta_T^{ -\frac{a}{2}} $, of which order decreases with the increasing tuning parameter $a$.
In this case, we select the smoothing parameter $\delta$ and the batch size $\Delta_T$ to minimize the regret defined in \eqref{eq:regret minimization strongly}. 
The accumulative estimation error is one of the dominant terms in the regret defined in \eqref{eq:regret minimization strongly} when $a \in (0,\frac{4}{3}]$ and becomes negligible when $a > \frac{4}{3}$, which results in a smaller regret order than in the convex case. 
\end{remark} 

\section{SIMULATION}\label{sec:simulation} 
\begin{figure}[htbp]
\begin{center}
\centerline{\includegraphics[width=0.9\columnwidth]{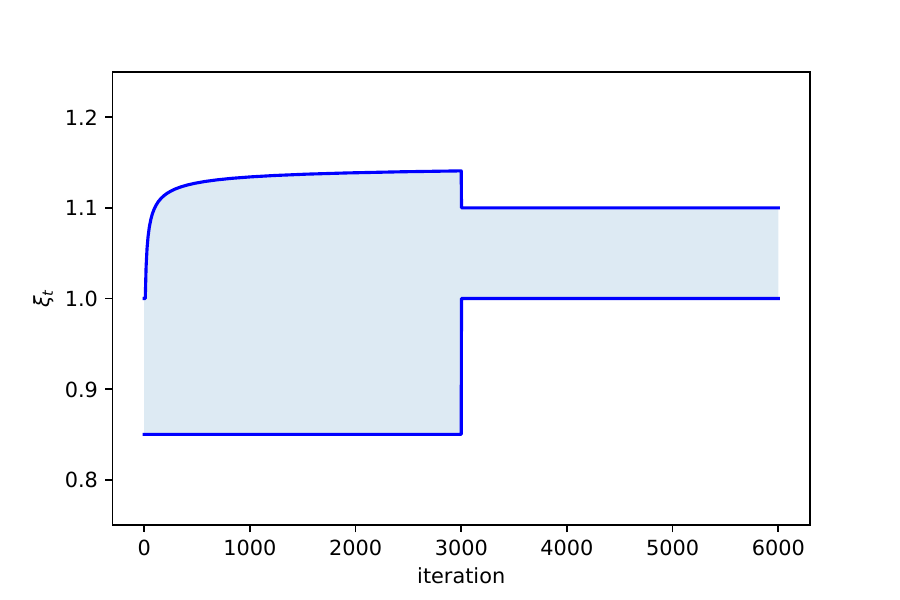}}
\caption{Distribution range of the uniform random variable $\xi_t$.}
\label{fig_rv}
\end{center}
\vskip -0.2in
\end{figure}

In this section, we consider the parking lot dynamic pricing problem, see \cite{ray2022decision}.  
Factors such as parking prices, availability, and locations generally influence driving decisions. This encourages us to dynamically adjust the parking price according to real-time demand. 
Denote $r_t \in [0,1]$ as curb occupancy rate. Let the occupancy rate be influenced by the price $x_t$ and environmental uncertainties $\xi_t$, which is 
\begin{equation*}
    r_t =\xi_t +A x_t,
\end{equation*}
where $A = -0.15$ is the estimated price elasticity, which is determined by \cite{ray2022decision} through analysis of the real-world data. The uncertainty $\xi_t $ is distributed according to the time-varying distribution $\DC_t$, which will change periodically according to environmental effects such as climate condition and dates.  
Specifically, to make it easy to find a parking space, it is desirable to maintain an occupancy rate of $70 \%$. 
Hence, the loss function is defined as 
\begin{equation*}
J(x_t,\xi_t)
     = \| \xi_t + A x_t -0.7 \|^2+\frac{\nu}{2}\|x_t\|^2, 
\end{equation*}
where $\nu = 0.001$ is the regularization parameter.  To avoid the overcrowded situation, we aim at minimizing the risk-averse objective function
\begin{equation}\label{eq:simulation cvar}
C_t(x_t)= \mathop{\cvar_\alpha}\limits_{\xi_t \sim \DC_t}[J(x_t,\xi_t)],
\end{equation}
where the risk level is selected as $\alpha = 0.5$. 
We assume that the random variable $\xi_t$ has a continuous uniform distribution and lies in the time-varying distribution range $[L_t,R_t]$, which is selected as
\begin{equation*}
    [L_t,R_t] = \left\{ \begin{array}{ll}
   [0.85, 1.15- 0.5t^{-0.5}]      & {\rm if}~t<T/2 \\
     \lbrack 0.85+0.5t^{-0.1}, 1.1 \rbrack    & {\rm if}~t\geq T/2.
    \end{array}\right.     
\end{equation*}
The time horizon is selected as $T =6000$. Fig.~\ref{fig_rv} depicts the distribution range of the random variable $\xi_t$. 

We use Algorithm~\ref{alg:algorithm} to update the parking prices in the non-stationary environment with changing distributions. We select the restarting period as $\Delta_T = 200$. We set the initial price value as $x_0 = 1$ and restrict the potential prices in the range of $[1,5]$. The smoothing parameter for the zeroth-order optimization is set as $\delta = 0.05$.  We set $a = \frac{2}{3}$ and $c=10$ for the sampling requirement \eqref{eq:sampling requirement}. Accordingly, we select $n_t = 8$ and run the algorithm for 10 trials. Shaded areas represent $\pm$ one standard deviation over 10 runs. 

\begin{figure}[htbp]
\begin{center}
\centerline{\includegraphics[width=0.9\columnwidth]{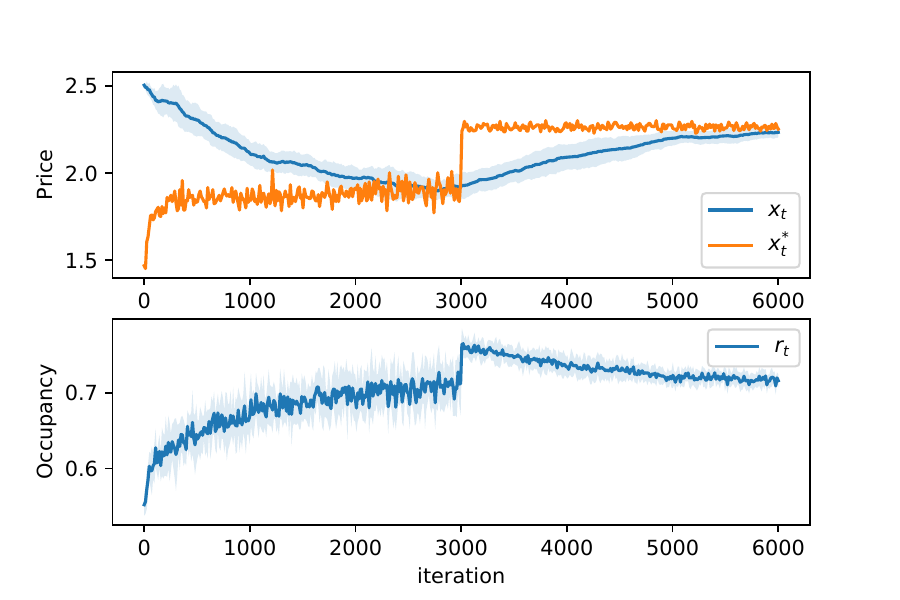}}
\caption{From top to bottom: the parking price $x_t$ under Algorithm~\ref{alg:algorithm} and the optimal parking price $x_t^{*}$; the resulted occupancy $r_t$ under Algorithm~\ref{alg:algorithm}. }
\label{fig_price}
\end{center}
\vskip -0.2in
\end{figure}

\begin{figure}[htbp]
\begin{center}
\centerline{\includegraphics[width=0.9\columnwidth]{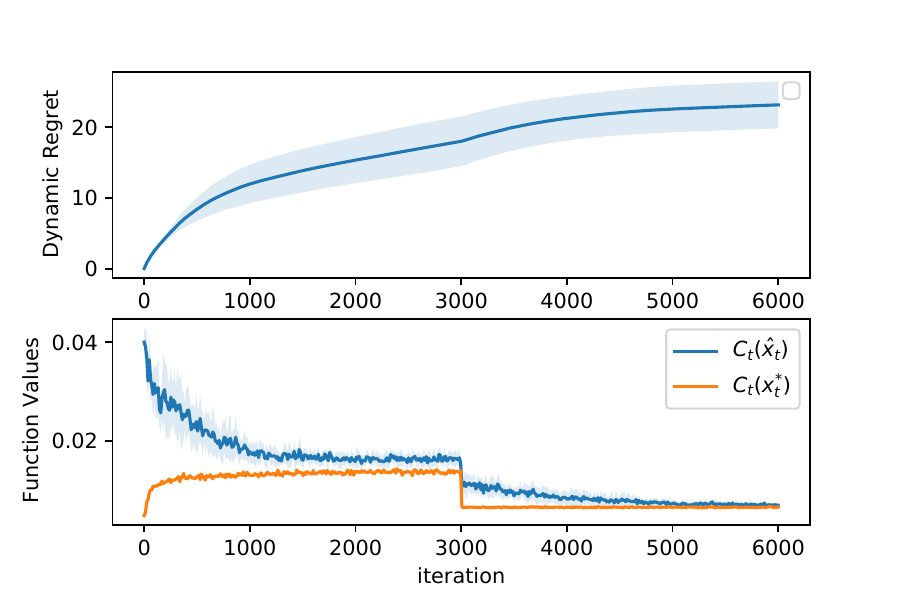}}
\caption{From top to bottom: dynamic regret; the CVaR values achieved by Algorithm~\ref{alg:algorithm} and the minimum CVaR values.}
\label{fig_dr}
\end{center}
\end{figure}

\begin{figure}[htbp]
\begin{center}
\centerline{\includegraphics[width=0.9\columnwidth]{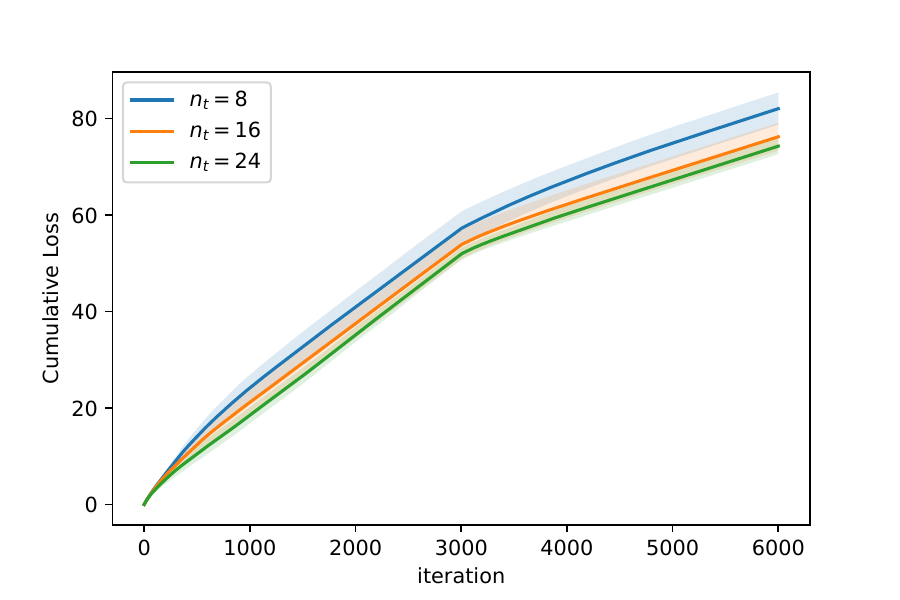}}
\caption{Accumulated loss achieved by Algorithm~\ref{alg:algorithm} under sampling strategies with constant number $n_t = 8,16,24$, respectively. }
\label{fig_CL}
\end{center}
\vskip -0.2in
\end{figure}

The simulation results of Algorithm~\ref{alg:algorithm} are presented in Figs.~\ref{fig_price}-\ref{fig_CL}. 
The top subfigure of Fig.~\ref{fig_price} depicts the parking price $x_t$ and the optimal price $x_t^{*}$, which minimizes the CVaR function defined in \eqref{eq:simulation cvar}. The bottom subfigure of Fig.~\ref{fig_price} depicts the resulting occupancy $r_t$. Since the CVaR function does not have a closed-form expression, we search for optimal prices by drawing a sufficiently large number of samples.  
More specifically, at each iteration, we select 100 distinct points uniformly in the continuous set $\mathcal{X}$ and identify the point that corresponds to the minimum CVaR values.  It can be observed that the price $x_t$ generated by Algorithm~\ref{alg:algorithm} catches up with the optimal parking price. 
Fig.~\ref{fig_dr} shows the CVaR value under the prices generated by Algorithm~\ref{alg:algorithm}, i.e.,  $C_t(\hat{x}_t)$
and the CVaR value under the optimal prices, i.e., $C_t(x_t^{*})$, and the dynamic regret ${\rm DR}(T)$.
Moreover, to explore the effect of different sampling numbers on the algorithm,
we use the sampling strategies with parameters $c = 10$ and $a \in \{\frac{2}{3},1, \frac{4}{3}\}$. The corresponding sampling number is $n_t\in\{8,16,24\}$, respectively. As the optimal prices $x_t^{*}$ as well as the minimum CVaR value $\cvar(x_t^{*})$ are the same in these three cases, we use the accumulated loss under the designed algorithm, i.e., $\sum_{t=1}^T C_t(\hat{x}_t) $ as the measure.   It is shown in Fig~\ref{fig_CL} that, more samples lead to a smaller cumulative loss, which illustrates our theoretical results. 



%

\section{CONCLUSIONS}\label{sec:conclusion}
In this paper, we investigated risk-averse learning with time-varying distributions. We employed a risk-averse learning algorithm that queries the function values for multiple times to estimate the gradient of CVaR and proved that the accumulative error of the CVaR gradient is bounded with high probability. By leveraging the restarting procedure, we bound the dynamic regret in terms of the distribution variations for both convex and strongly convex cases. The theoretical results suggest that the increasing sampling numbers reduce the order bound achieved by the designed algorithm, and this reduction ceases after the total sampling number reaches a certain threshold. Moreover, the strongly convex assumption leads to a smaller regret order bound than in the convex case. 

\section{Appendix}\label{sec:appendix}
\textit{Proof of Lemma~\ref{lemma:cva-wasserstein}:} Define the augmented functions 
\begin{align*}
    &L_{\DC_X}(v) = v+\frac{1}{\alpha} \mathbb{E}_{X \sim \DC_X}[f(X)-v]_{+} ,\\
    &L_{\DC_Y}(v) = v+\frac{1}{\alpha} \mathbb{E}_{Y \sim \DC_Y}[f(Y)-v]_{+},
\end{align*}
where $\DC_X$ and $\DC_Y$ are the distributions of the random variables $X$ and $Y$, respectively. According to \cite{rockafellar2000optimization}, we have
\begin{align*}
&    \mathop{\cvar_\alpha}\limits_{  X \sim \DC_X }[f(X)] =  \min\limits_{v}L_{\DC_X}(v), \\
 &   \mathop{\cvar_\alpha}\limits_{  Y \sim \DC_Y }[f(Y)] = \min\limits_{v}L_{\DC_Y}(v).
\end{align*}
We assume that $v_x = \argmin_{v}L_{\DC_X}(v)$ and $v_y = \argmin_{v}L_{\DC_Y}(v)$. Then, we have
\begin{equation*}
    L_{\DC_X}(v_x) = \mathop{\cvar_\alpha}\limits_{X \sim \DC_X}[f(X)], \quad 
     L_{\DC_Y}(v_y) = \mathop{\cvar_\alpha}\limits_{Y \sim \DC_Y}[f(Y)].
\end{equation*}
It follows that
\begin{align*}
  & \hspace{1em}\mathop{\cvar_\alpha}\limits_{X \sim \DC_X}[f(X)]  -  \mathop{\cvar_\alpha}\limits_{Y \sim \DC_Y}[f(Y)] \nonumber \\
  &= L_{\DC_X}(v_x) -L_{\DC_Y}(v_y) \nonumber \\
  &\leq  L_{\DC_X}(v_y) - L_{\DC_Y}(v_y) \nonumber \\ 
  &= v_y + \frac{1}{\alpha}\mathbb{E}_{X\sim \DC_X}[f(X)-v_y]_{+}  \nonumber \\
  &\hspace{1em} -v_y - \frac{1}{\alpha}\mathbb{E}_{Y\sim \DC_Y}[f(Y)-v_y]_+ \nonumber \\
 &=  \frac{1}{\alpha} \mathbb{E}_{X\sim \DC_X}[g(X)] - \frac{1}{\alpha} \mathbb{E}_{Y\sim \DC_Y}[g(Y)],
\end{align*}
where we define $g(x) = [f(x)-v_y]_+ $.
The first inequality is due to the fact that $v_x = \argmin_{v}L_{\DC_X}(v)$.
According to the definition of $g(x)$, we have
\begin{align}\label{eq:g:Lips}
    |g(x) - g(y)| &=  [f(x)-v_y]_+ - [f(y)-v_y]_+ \nonumber \\
    &\leq  [f(x)-f(y)]_+  \nonumber \\ 
    &\leq   [f(x)-f(y)] \le L_0 \left\| x-y\right\|,
\end{align}
where the first inequality follows from the fact that $a_+ - b_+ \leq [a-b]_+$, for $\forall a,b \in \mathbb{R}$.
From \eqref{eq:g:Lips}, we conclude that the function $g(x)$ is $L_0$-Lipschitz continuous. Hence,
\begin{align*}
     &\hspace{1em}\mathop{\cvar_\alpha}\limits_{X \sim \DC_X}[f(X)]  -  \mathop{\cvar_\alpha}\limits_{Y \sim \DC_Y}[f(Y)]  \\
& \leq  \frac{1}{\alpha}   \mathbb{E}_{X\sim \DC_X}[g(X)]- \frac{1}{\alpha} \mathbb{E}_{Y\sim \DC_Y}[g(Y)]  \\
& \leq  \frac{L_0}{\alpha}  W_1(\DC_X,\DC_Y),
\end{align*}
where the last inequality follows from the Kantorovich-Rubinstein Duality of the Wasserstein distance, see \cite{edwards2011kantorovich}.

Following similar arguments, we can obtain the other side of the inequality.   
Here completes the proof. \hfill $\qed$

\textit{Proof of Lemma~\ref{lemma:strongconvex}:}
1)
We define the function $h(x,\xi_t)=J(x,\xi_t) - \frac{m}{2}\left\| x\right\|^2$. As $J(x,\xi_t)$ is $m$-strongly convex in $x$ for every $\xi_t \in \Xi$, we have that $h(x,\xi_t)$ is convex in $x$ for every $\xi_t \in \Xi$. Using Lemma \ref{lemma:cvar:convex}, we have that $\mathrm{CVaR}_{\alpha}\left[ h(x,\xi_t)\right]$ is convex in $x$. According to the translation invariance property of CVaR, we obtain 
\begin{align*}
    \mathrm{CVaR}_{\alpha}\left[ h(x,\xi_t)\right] &= \mathrm{CVaR}_{\alpha}\left[ J(x,\xi_t)\right] - \frac{m}{2}\left\| x\right\|^2 \nonumber \\
   &=  C_t(x) - \frac{m}{2}\left\| x\right\|^2.
\end{align*}
Using trivial arguments in \cite{boyd2004convex}, we conclude that $C_t(x)$ is $m$-strongly convex in $x$.

2) As shown in Lemma 2.8 in \cite{hazan2016introduction}, if the original function is strongly convex, the smoothed function is also strongly convex. The proof is complete. \hfill $\qed$

\begin{lemma}
 \label{lemma:function variation}
Consider the function $C_t^\delta(x): \XC \rightarrow \mathbb{R}$, define the function sequences over iterations horizon $T$ as $\{C_t^\delta(x_t)\}_{t=1}^{T}$,  we have that 
\begin{align}\label{eq:loss-function variation}
   & \sum_{j=1}^s   \sum\limits_{t \in \TC_j}  \bigg(C_t^\delta(x_j^{\delta,\ast}) -  C_t^\delta(x_t^{\delta,\ast})\bigg)  \nonumber \\ 
   &\leq  \frac{2 L_0\Delta_T}{\alpha} \sum_{t=2}^T W_1(
   \DC_t, \DC_{t-1}
   ).
\end{align}  
\end{lemma}
\textit{Proof of Lemma~\ref{lemma:function variation}:} This proof is adopted from Proposition 2 of \cite{besbes2015non}. 
First we denote  $$V_j = \sum_{t \in \TC_j} \sup_{x \in \XC^\delta}|C_t^\delta(x)-C_{t-1}^\delta(x)|$$ as the function variation over batch $\TC_j$, it is straightforward to write $    \sum_{j=1}^s V_j    = \sum_{t=2}^T \sup_{x \in \XC^\delta}|C_t^\delta(x) - C_{t-1}^\delta(x)|$.

Let $\bar{\tau}_j$ be the first epoch of batch $\TC_j$, for $j=1,\dots, s$, we have 
\begin{align*}\label{eq:variation bound1}
& \hspace{1em}    \sum_{t \in \TC_j} C_t^\delta(x_j^{\delta,\ast})  -\sum_{t \in \TC_j} C_t^\delta(x_t^{\delta,\ast}) \nonumber \\
&\leq \sum_{t \in \TC_j}C_t^\delta(x_{\bar{\tau}_j}^{\delta,\ast})-C_t^\delta(x_t^{\delta,\ast}) \nonumber \\ 
&\leq \Delta_T \cdot \max_{t \in \TC_j}\{C_t^\delta(x_{\bar{\tau}_j}^{\delta,\ast})-C_t^\delta(x_t^{\delta,\ast})\}.
\end{align*}
In the following we will prove   $\max_{t \in \TC_j} \{C_t^\delta (x_{\bar{\tau}_j}^{\delta,\ast} )-C_t^\delta(x_t^{\delta,\ast}) \} \leq 2 V_j$ by contraction. Suppose otherwise, 
there exists an iteration $\tilde{t} \in \TC_j$ such that $C_{\tilde{t}}^\delta (x_{\bar{\tau}_j}^{\delta,\ast})-C_{\tilde{t}}^\delta(x_{\tilde{t}}^{\delta,\ast})>2 V_j$. It implies that 
\begin{equation*}
C_t^\delta(x_{\tilde{t}}^{\delta,\ast}) \leq C_{\tilde{t}}^\delta(x_{\tilde{t}}^{\delta,\ast})+V_j<C_{\tilde{t}}^\delta(x_{\bar{\tau}_j}^{\delta,\ast})-V_j \le C_t^\delta(x_{\bar{\tau}_j}^{\delta,\ast}),
\end{equation*}
where the first and the last inequality results from the fact that $V_j$ is the maximal variation over batch $\TC_j$. 
Hence, we have  $$  \sum_{t \in \TC_j} C_t^\delta(x_j^{\delta,\ast})  -\sum_{t \in \TC_j} C_t^\delta(x_t^{\delta,\ast}) \leq 2 \Delta_T V_j.$$ 
Summarize the variation along batches $\{\TC_1,\dots, \TC_s \}$,  it results 
\begin{align}\label{eq:function variation proof}
 & \hspace{1em} \sum_{j=1}^s   \bigg(  \sum_{t \in \TC_j} C_t^\delta(x_j^{\delta,\ast})  -\sum_{t \in \TC_j} C_t^\delta(x_t^{\delta,\ast})\bigg)  \nonumber \\
    &\leq \sum_{j=1}^s 2 \Delta_T V_j
    \le  \frac{2 L_0}{\alpha}\Delta_T \sum_{t=2}^T W_1(
   \DC_t, \DC_{t-1}
   ).
\end{align}
Here is the proof. \hfill $\qed$
\addtolength{\textheight}{-12cm}   %




\bibliographystyle{unsrt}        
\bibliography{autosam}           

\end{document}